\documentstyle[12pt,fleqn]{article}
\def\ga{\mathrel{\mathpalette\fun >}}
\def\fun#1#2{\lower3.6pt\vbox{\baselineskip0pt\lineskip.9pt

\ialign{$\mathsurround=0pt#1\hfill##\hfil$\crcr#2\crcr\sim\crcr}}}
\begin{document}
\begin{titlepage}
\null\vspace{-72pt}
\begin{flushright}
{\footnotesize
FERMILAB--Pub--96/88-A\\
OSU-TA-??/96\\
astro-ph/9604444\\
April 1996 \\
Submitted to {\em Phys. Lett. B }}
\end{flushright}
\renewcommand{\thefootnote}{\fnsymbol{footnote}}
\vspace{0.15in}
\baselineskip=24pt

\begin{center}
{\Large \bf  Non-equilibrium symmetry restoration\\
 beyond one loop }\\
\baselineskip=14pt
\vspace{0.75cm}

\vspace{0.3cm}
Antonio Riotto\footnote{Electronic address:
{\tt riotto@fnas01.fnal.gov}}\\
\vspace{0.2cm}
{\em NASA/Fermilab Astrophysics Center\\
Fermi National Accelerator Laboratory, Batavia, Illinois~~60510}\\
\vspace{0.3cm}
and\\
\vspace{0.3cm}
Igor I. Tkachev\footnote{Electronic address: {\tt  
tkachev@wollaston.mps.ohio-state.edu}}\\
\vspace{0.2cm}
{\em Department of Physics, The Ohio State University, Columbus, OH~~  
43210\\ 

and\\
Institute for Nuclear Research of the Academy of Sciences of Russia\\
Moscow~~117312, Russia}\\
\end{center}

\baselineskip=24pt

\begin{quote}
\hspace*{2em}  We calculate the strength of symmetry restoration
effects in highly non-equilibrium states which can arise, for example, during
preheating after inflation.  We show that in certain parameter range 
the one-loop results are unstable, requiring summation of  multiloop
diagrams. We solve this problem for the $O(N)$ model in the large $N$-limit
and show that the symmetry restoration may be less effective than what predicted by the one-loop estimate.
\vspace*{8pt}


\renewcommand{\thefootnote}{\arabic{footnote}}
\addtocounter{footnote}{-2}
\end{quote}
\end{titlepage}

\newpage

\baselineskip=24pt
\renewcommand{\baselinestretch}{1.5}
\footnotesep=14pt


\def\LHS{{\sc lhs}}
\def\RHS{{\sc rhs}}
\def\GUT{{\sc gut}}
\def\LTE{{\sc lte}}
\def\VEV{{\sc vev}}
\def\beq{\begin{equation}}
\def\eeq{\end{equation}}
\def\beqa{\begin{eqnarray}}
\def\eeqa{\end{eqnarray}}
\def\tr{{\rm tr}}
\def\ph{\tilde{\gamma}}
\def\g{\tilde{g}}
\def\x{{\bf x}}
\def\p{{\bf p}}
\def\k{{\bf k}}
\def\z{{\bf z}}
\def\re#1{{[\ref{#1}]}}
\def\eqr#1{{Eq.\ (\ref{#1})}}
\def\trh{{T_{\rm RH}}}
\def\tph{{T_{\rm PH}}}

It has been recently realized \re{explosive} that the process of  
reheating after the inflationary era \re{linde} may consist of different  
epochs. 

In the first stage, the effects of stimulated decays and annihilation may lead 
to an extremely effective explosive particle
production. 
The energy is released in the form of inflaton decay
products, whose occupation numbers are extremely large, and have
energies much smaller than the temperature that would have been
obtained by an instantaneous conversion of the inflaton energy  
density into equilibrated radiation. 

Since it requires several scattering times for the low-energy decay
products to form a thermal distribution, it is rather reasonable to
consider the period in which most of the energy density of the
Universe was in the form of non-thermal quanta produced by
inflaton decay as a separate cosmological era.  Following \re{explosive}
we refer to the final stages of parametric resonance as the {\it preheating} 
epoch  to distinguish it from the subsequent stages of particle decay and  
semiclassical thermalization \re{st} when the inflaton decay can be still efficient,
but amplitudes of fluctuating fields became gradually smaller 
because of Universe expansion.

Several aspects of the theory of explosive reheating have been  
studied in the case of slow-roll inflation \re{noneq} and first-order  
inflation
\re{kolb}. In particular, the phenomenon of symmetry restoration  
during the preheating era has
been investigated recently by Kofman,
Linde, and Starobinski \re{KLSSR} and by Tkachev \re{tkachev} in the  
framework of typical chaotic
inflationary models.  It was shown that nonthermal symmetry  
restoration processes
during the nonequilibrium stage of preheating may be very efficient
with important implications for models containing relatively low scales
like the invisible axion model or supersymmetry with Polonyi fields, and may be for 
Grand Unified Theories (GUT).  For example, if a \GUT\ symmetry is restored during the  
preheating
epoch, the subsequent symmetry breaking phase transition will
reintroduce the problems of monopoles \re{mono} or domain walls
\re{dw}. In any case, if one askes whether particular high-energy 
symmetries are restored after inflation or not, one has to study 
the non-equilibrium preheating epoch.

We consider a model with the following potential
\begin{equation}
\label{pot}
V_0(\phi, \eta)=\lambda{\phi}^4 
+\frac{\alpha}{4!}\left(\eta^2-\eta_0^2 \right)^2 +
\frac{g}{2}\phi^2\eta^2\, . 
\end{equation}
The field $\phi$ corresponds to the inflaton. While inflaton self-coupling 
is restricted to be
$\lambda \approx 10^{-13}$, there is no such severe restrictions on other
couplings. Among products of the inflaton decay there 
can be different species, but for simplicity we assume here that 
quanta of the field $\eta$ constitute the major channel of its 
decay, which means in particular that
$g \gg \lambda$. We assume also that $\eta$ transforms as a vector under the 
action of $O(N)$,  {\it i.e.} $\eta^2 = \eta_a \eta^a$ with $a=1,\cdot\cdot\cdot,N$. 

In previous papers \re{explosive},\re{KLSSR},\re{tkachev} the effects of 
the coupling
$\alpha$ was neglected, which corresponds, in fact,  to the one loop
approximation. The importance of multiloop correction was noticed in Ref. 
\re{tkachev}. We shall show here that they change the results dramatically 
when $\alpha \gg g$ and the one-loop approximation actually breaks down
in this region.

We do not consider an implications of non-zero $\alpha$ on parametric
resonance itself, but consider the effects of symmetry restoration
only. We assume that at the end of the resonance stage the energy density
which builds up in the fluctuations of field $\eta$, and which we denote as 
$\rho_\eta$, is some fraction of 
inflaton energy density. Simple assumption would be that this fraction
equals to one half \re{explosive}. Numerical integration
of Ref. \re{st} for the case of simplest 
$\lambda \phi^4$ model confirms that the resonance ends 
when half of 
the energy is in fluctuations, however it might not be true for $g \gg \lambda$. 
Morover, because of
the expansion of the Universe during preheating, $\rho_\eta$
is many orders of magnitude smaller than the initial inflaton energy density.
Since this is
model dependent, at present we keep $\rho_\eta$ as a free parameter.

We can parametrize the distribution function of the created quanta at 
preheating as 
$f \equiv f(k/k_*)$ where $k$ is particle momenta. What is important 
here is the smallness of typical particle momenta, $k_*$, with respect to
the temperature that one would get in the case of istantaneous reheating  neglecting the expansion of the Universe.
In the extreme case when all inflaton couplings are not very different, 
$k_*$ is of order of inflaton mass, $k_* \sim 10^{13}$ GeV.
For definitness, let us assume that the distribution function is of the form  
\begin{equation}
f(k)=A\:\delta\left(|{\bf k}|-|{\bf k}_*|\right).
\label{ind}
\end{equation}
Parametrically, and that is what we are interested in here, Eq. (\ref{ind}) 
will reproduce the correct unswer. Moreover, the distribution is really 
peaked at the end of the resonance stage in reasonably wide range of model
parameters.

The constant $A$ (or its initial value in our problem) can be fixed by computing
the energy  density of $\eta$ particles
and setting it equal to $\rho_\eta$.  
Notice that, since the order parameter changes in time, {\it e.g.} during the 
process of symmetry restoration, the energy of particles coupled to it 
changes too, but  
three-momenta keep constant in an homogeneous background. Moreover,
in this case the effective potential simply coincides with the energy density,
which simplifies the calculations. To answer the question whether the symmetry
tends to be restored or not, it is sufficient to consider homogeneous
background only.

As opposed to large-angle scattering processes, forward-scattering
processes do not alter the distribution function of the particles
traversing a gas of quanta, but simply modify the dispersion
relation. This remains true also in the case of a nonequilibrium
system. Forward scattering is manifest, for example, as ensemble
and scalar background corrections to the particle masses.  Since the
forward scattering rate is usually larger than the large-angle
scattering rate responsible for establishing a thermal distribution,
the nonequilibrium ensemble and scalar background corrections are
present even before the initial distribution function 
relaxes to its thermal value.  These considerations allow us to  
impose
$k^0=\sqrt{|{\bf k}|^2+m_\eta^2(\bar{\eta})}$ as the dispersion relation  
for the particles created by the parametric resonance, where $\bar{\eta}$
is the order parameter, with $\bar{\eta}=\eta_0$ in vacuum, and 
$m_\eta^2(\bar{\eta})\delta_{ab}={\alpha}\left[(\bar{\eta}^2-\eta_0^2)\delta_{ab}+
2\bar{\eta}_a\bar{\eta}_b\right]/3!$.

Let us first consider the case $N=1$. 
We can not use the imaginary-time formalism to determine the  
effective
potential for the scalar field $\bar{\eta}$ during the nonequilibrium
preheating period since in the nonequilibrium case there is no
relation between the density matrix of the system and the time
evolution operator, which is of essential importance in the
formalism. There is, however, the real-time formalism of Thermo Field
Dynamics (TFD), which suites our purposes \re{tfd}. This approach  
leads to a $2\times 2$ matrix structure for the free propagator (only  
the (11)-component is physical)
\begin{eqnarray}
\left(\begin{array}{cc}
D_{11}(K) & D_{12}(K)\\
D_{21}(K) & D_{22}(K)
\end{array}\right)&=&
\left(\begin{array}{cc}
\Delta(K) & 0\\
0 & \Delta^*(K)
\end{array}
\right)
+\left(\begin{array}{cc}
f(k) & \theta(k_0)+f(k)\\
\theta(-k_0)+f(k) & f(k)
\end{array}
\right)\nonumber\\
&\times &\:2\pi \delta[K^2-m_\eta^2(\bar{\eta})],
\end{eqnarray} 
with the usual vacuum Feynman propagator
\begin{equation}
\Delta(K)=\frac{i}{K^2-m_\eta^2(\bar{\eta})+i\epsilon}.
\end{equation}
The one-loop effective potential $V_1(\bar{\eta})$ can be easily obtained  
solving
the equation 
\begin{equation}
\frac{d V_1}{d m^2_\eta}={\cal T}_1=\frac{1}{2}\int\:\frac{d^4  
K}{(2 \pi)^4}\:
D_{11}(K)\equiv\frac{1}{2}\left(I_0+I_f\right) ,
\end{equation}
where ${\cal T}_1$ is the one-loop tadpole diagram which receives  
contribution only from the diagram in Fig. 1 (since the final  
external leg is fixed to be that of the physical field) and we have  
separated the usual zero-temperature contribution from the one given  
by the gas of $\eta$-quanta with distribution function $f$, Eq.  
(2). One then  obtains 
\begin{equation}
V_1(\bar{\eta})=V_1^{(0)}+V_1^{(f)}=V_1^{(0)}+\frac{A\:|{\bf k}_*|^2}{4\pi^2}
\sqrt{|{\bf k}_*|^2+m_\eta^2(\bar{\eta})},
\end{equation}
where $V_1^{(0)}$ is the one-loop zero-temperature effective potential.
With the distribution function Eq. (\ref{ind}), the  particle number density is
\begin{equation}
n=\frac{1}{(2\pi)^3}\int d^3k \:f(k)=\frac{A\:|{\bf k}_*|^2}{2\pi^2},
\label{pnd}
\end{equation}
and the particle-dependent part of the effective potential reads
$V_1^{(f)}={n}\sqrt{|{\bf k}_*|^2+m_\eta^2(\bar{\eta})}/2$. We would like to
mention for the future use that Eq. (\ref{pnd}) is valid
only  in the  one-loop approximation .
Expanding  
$V_1^{(f)}$ around $\bar{\eta}=0$, we obtain that the coefficient of  
$\bar{\eta}^2$-term in the potential (or the effective mass of the field) 
is changed by interaction with the medium by an amount 
$\alpha n/|{\bf k}_*|\sim \alpha\rho_{\eta}/|{\bf k}_*|^2$ \re{KLSSR}, \re{tkachev}. The  
presence of nonthermalized $\eta$-quanta may lead to symmetry  
restoration if $\rho_{\eta}/|{\bf k}_*|^2 \ga \eta_0^2$.

Our main concern in this Letter is on what happens when multi-loop  
corrections are included in the computation of the nonequilibrium effective  
potential. We will show that in certain range of parameters 
the effective potential gets large contributions at  
the  two-loop level and higher orders of perturbation theory, even larger than the one-loop contribution. Perturbation  theory is then  lost unless a proper resummation
can be done. We give an example in which the resummation can be performed 
exactly.

Let us first consider the theory described by the potential in  
\eqr{pot}. By applying the rules of TFD we obtain the two-loop  
tadpole diagrams
as drawn in Fig. 2 (plus counter-terms arising at one-loop).  
Underlined diagrams are identically zero since they contain
$\delta(K^2-m^2)\delta((K-P)^2-m^2)\delta(P^2-m^2)=0$. 


We first consider the simplest case of  diagram 2a
\begin{equation}
\frac{dV_{(2{\rm a})}}{d  
m^2_\eta}=\frac{i\:\alpha}{4}\:\int\:\frac{d^4 K}{(2  
\pi)^4}\:\int\:\frac{d^4 P}{(2 \pi)^4}\:\left[D^2_{11}(P)\:D_{11}(K)-
D_{12}(P)\:D_{21}(P)\:D_{22}(K)\right].
\end{equation}
Observing that the tadpole does not have any imaginary part and that  
${\rm Re}\:D_{11}(K)={\rm Re}\:D_{22}(K)$, one can factorize the  
$K$-integral out and, using the mass-derivative formula \re{mass},  
obtain
\begin{eqnarray}
\label{2a}
\frac{dV_{(2{\rm a})}}{d  
m^2_\eta}&=&\frac{i\:\alpha}{4}\:\int\:\frac{d^4 K}{(2  
\pi)^4}\:{\rm Re}\:D_{11}(K)\:
\frac{d^4 P}{(2 \pi)^4}\:\left[\Delta^2(P)+f\left(\Delta^2(P)-
(\Delta^{*}(P))^2\right)\right]\nonumber\\
&=&\frac{i\:\alpha}{4}\:\int\:\frac{d^4 K}{(2 \pi)^4}\:{\rm  
Re}\:D_{11}(K)\:
\left[\int\:\frac{d^4 P}{(2 \pi)^4}\:
\Delta^2(P)-i\:\frac{\partial I_f}{\partial  
m^2_\eta}\right]\nonumber\\
&=&\frac{i\:\alpha}{4}\:{\rm  
Re}\left(I_0+I_f\right)\left(J_0-i\:\frac{\partial I_f}{\partial  
m^2_\eta}\right),
\end{eqnarray}
where we have defined  $J_0=(2 \pi)^{-4}\:\int d^4 P\:\Delta^2(P)$.  
\eqr{2a} shows the absence of pinch sungularities, or  
$\Delta(P)\Delta^*(P)$ terms in the final result. The cancellation  
procedure works essentialy in the same way as in the equilibrium case  
and is totally independent of the distribution $f$ \re{pert}.
After renormalization, the term proportional to $I_f\:J_0$  gives a  
contribution $\sim \alpha\:I_f\:{\rm ln}\:m^2_\eta$ which is  
suppressed by ${\cal O}(\alpha)$ with respect to the one-loop  
result. One should not claim victory to soon, though. Let us extract  
the  $f$-dependent part only: its    contribution goes like  
$\alpha\:I_f\:\left(\partial I_f/\partial m^2_\eta\right)$.  
Expanding such an expression around $\bar{\eta}=0$, we see that the  
contribution of diagrams of Fig. 2a is of order of  
$(\alpha\:\rho_\eta/|{\bf k}_*|^4)$ with respect to the one-loop result.  
Since $\alpha$ may be  $\sim 1$ and $|{\bf k}_*|^4/\rho_\eta\ll 1$ at the  
preheating era, we discover  that the contribution from the two-loop  
tadpole diagrams 
is much larger than the one-loop result by several orders of  
magnitude  and perturbation theory is lost!.

Things get even worse when we consider  $p$-loop diagrams. The  
tadpole diagrams shown in Fig. 3 ($p\geq 2$) contribute
\begin{equation}
\frac{dV_{(3)}}{d  
m^2_\eta}=\frac{1}{2}\left(\frac{i\:\alpha}{2}\right)^{p-1}\:
{\rm Re}\left(I_0+I_f\right)\:\left(J_0-i\:\frac{\partial  
I_f}{\partial m^2_\eta}\right)^{p-1}.
\end{equation}
All the terms which contain a vacuum contribution are ultraviolet  
divergent and by cancellation with the counter-terms we are left with  
only the $f$-dependent part which has the following behaviour (with  
respect to the one-loop result)
\begin{equation}
\left(\alpha\:\frac{\partial I_f}{\partial  
m^2_\eta}\right)_{\phi=0}^{p-1}\sim  
\left(\frac{\alpha\:\rho_\eta}{|{\bf k}_*|^4}\right)^{\frac{p-1}{2}}\gg  
1.
\end{equation}

It is possible to add the tadpoles in a different way. Let us  
consider the $p$-loop contributions shown in Fig. 4, the so-called  
daisy diagrams. Using again the mass-derivative formula, we obtain  
the expression for a product of $p\geq 2$ successive propagators
\begin{equation}
\label{xx}
\frac{dV_{(4)}}{d  
m^2_\eta}=\frac{1}{2}\:\left(\frac{\alpha}{2}\right)^{p}\:\left(I_ 
0+I_f
\right)^p\:\frac{1}{p!}\:\left(\frac{\partial}{\partial  
m^2_\eta}\right)^p\left(I_0+I_f\right).
\end{equation}
This expression has also to be renormalized. For $p\ge2$ the  
ultraviolet singularities are resummed in $I_0$ and are cancelled by  
adding a series of counter-terms arising at one-loop. The result is 
\begin{equation}
\frac{dV_{(4)}}{d m^2_\eta}+{\rm  
C.T.}=\frac{1}{2}\:\left(\frac{\alpha}{2}\right)^{p}\:I_f^N\:\frac{1} 
{p !}\:\left(\frac{\partial}{\partial  
m^2_\eta}\right)^p\:\left(I_0+I_f\right).
\end{equation}
It is clear that successive derivatives with respect to the squared  
mass give increasing inverse powers of the squared mass. The  
derivatives of $I_0$ give a subleading contribution with respect to  
the $f$-dependent part. Extracting  the $f$-dependent part and  
expanding such it around $\bar{\eta}=0$, we discover  that the contribution  
of the $p$-loop daisy diagrams  is again of order of  
$(\alpha\:\rho_\eta/|{\bf k}_*|^4)^{p}\gg 1$ with respect to the one-loop  
result.

The discussion above shows that in order to obtain a more accurate  
information about the issue of nonthermal symmetry restoration one  
must study an infinite series of diagrams in perturbation theory.  
This is exactly analogous to what happens in a simple  
$\lambda{\varphi}^4$ theory in equilibrium at finite temperature  
where the leading contributions to the effective potential in the  
infrared region come from the daisy and superdaisy multiloop graphs  
\re{dj}. 

To deal with this problem we need a self-consistent loop expansion of  
the effective potential in terms of the {\it full} propagator. Such a  
technique was developed some time ago by Corwall, Jackiw and  
Tomboulis (CJT) in their effective action formalism for composite  
operators\re{cjt} and may be also applied to nonequilibrium phenomena  
\re{chou}. One considers a generalization $\Gamma[\bar{\eta},G]$ of the  
usual effective action, which depends not only on $\bar{\eta}(x)$, but also  
on $G(x,y)$, a possible expectation value of the  time-ordered  
product $\langle T\:\eta(x)\:\eta(y)\rangle$. The  
physical solutions satisfy the stationary requirements
\begin{eqnarray} 
\label{gg}
\frac{\delta\Gamma[\bar{\eta},G]}{\delta\bar{\eta}(x)}&=& 0,\nonumber\\
\frac{\delta\Gamma[\bar{\eta},G]}{\delta G(x,y)}&=& 0.
\end{eqnarray}
The conventional effective action $\Gamma[\bar{\eta}]$ is given by  
$\Gamma[\bar{\eta},G]$
at the solution $G_0(\bar{\eta})$ of \eqr{gg}.
In this formalism it is possible to sum a large class of ordinary  
perturbation-series diagrams that contribute to the effective action  
$\Gamma[\bar{\eta}]$, and the gap equation which determines the form of the  
full propagator is obtained by a variational technique.

We now apply the CJT formalism in the limit of large  
$N$ when 
the next-to-leading terms can be exactly summed. 
In each order we keep only the term dominant in $N$ for large values  
of $N$. This allows us to resum exactly the series of the leading  
multiloop diagrams and to solve the gap equation for the full  
propagator without any approximation.

In order to obtain a series expansion of the effective action, one  
introduces the functional operator
\begin{equation}
D^{-1}_{ab}(\bar{\eta},x,y)=\frac{\delta^2\:I}{\delta\bar{\eta}_a(x)\delta
\bar{\eta}_b(y )},
\end{equation}
where $I$ is the classical action. The required series obtained by  
CJT is then \re{cjt}
\begin{equation}
\Gamma[\bar{\eta},G]=I(\bar{\eta})+\frac{1}{2}\:{\rm Tr}\:{\rm ln} \:D_0\:G^{-1}
+ \frac{1}{2}\:{\rm Tr}\:\left[D^{-1}\:G-1\right]+\Gamma_2[\bar{\eta},G],
\end{equation}
where  
$D_0^{-1}=-\left(\partial_\mu\partial^\mu+m^2\right)\delta_{ab}\delta^ 
4(x,y)$, and $\Gamma_2[\bar{\eta},G]$   
is a sum of all   
two-particle-irreducible vacuum graphs in the theory with vertices  
defined by the classical action with shifted fields $I[\bar{\eta}+\eta]$ and propagators set equal to  
$G(x,y)$. 


Previous calculations show that   among the multiloop graphs  
contributing to the effective potential in the $O(N)$-theory, only  
the daisy and superdaisy diagrams survive in the limit of large $N$  
\re{limit}. This enables us to consider in $\Gamma_2[\bar{\eta},G]$ only  
the 
graph of ${\cal O}(\alpha)$ given in Fig. 5. This is  the  
Hartree-Fock approximation which is known to be exact in the  
many-body version of our large $N$ limit.

It is more convenient to concentrate on the effective masses rather  
than on the effective potential.
By stationarizing the effective action $\Gamma[\bar{\eta},G]$ with respect  
to $G_{ab}$, we  obtain the gap equation
\begin{equation}
\label{p}
G_{ab}^{-1}(x,y)=D_{ab}^{-1}(x,y)+\frac{\alpha}{6}\left[\delta_{ab 
}\:G_{cc}(x,x)+2\:G_{ab}(x,x)\right]\:\delta^4(x,y).
\end{equation}
This equation is exact in the limit of large $N$ and contains all the  
informations about the dominant $N$-contributions to the full  
propagator. Indeed, the exact Schwinger-Dyson  equation reduces to  
\eqr{p} for large $N$. 
We Fourier-transform \eqr{p} and take $\bar{\eta}_a=0$. 
The gap equation then reads \re{cjt}
\begin{equation}
M^2=m^2+\frac{\alpha}{6}\:I_f(M^2)=
m^2+\frac{\alpha\:N\:A}{24\pi^2}
\frac{|{\bf k}_*|^2}
{\sqrt{|{\bf k}_{*}|^2+M^2}} \, .
\label{ge}
\end{equation}
We see that the one-loop results are stable when $|{\bf k}_*|^2\gg M^2$,
which translates to the condition $\rho_\eta \ll |{\bf k}_*|^4/\alpha$, where 
now $\rho_\eta$ indicates the total energy of the noninteracting gas
(the sum over all $\eta_a$).
In the opposite case the gap equation
is approximately solved by $M^3 \simeq \alpha\:N\: A\: |{\bf k}_*|^2 $.
Using $\rho_\eta \sim \alpha\:N^2 \:A^2 \:|{\bf k}_*|^4/ M^2$ we find
\begin{equation}
M^2 \simeq \sqrt{\alpha \rho_\eta} \, .
\label{mge}
\end{equation}
In other words, the strength of symmetry restoration measured
in terms of the effective temperature $T^2_{\rm eff}/12\equiv <(\eta-\bar{\eta})^2>$
is given by $T^2_{\rm eff}/12 \approx
M^2/\alpha \approx \sqrt{\rho_\eta/\alpha}$. This result and  Eq. (\ref{mge}) 
are intuitively understandable. Indeed, in this regime the contribution
to the energy from self-coupling is important and $\rho_\eta$ is saturated by 
the self-interaction term in Eq. (1).
This can be obtained in Hartree-Fock approximation directly applied to
equations of motion for the $\eta$-field (in the large $N$-limit Hartree-Fock
approximation becomes exact).

To summarize. In the limit
$ |{\bf k}_*|^4 \gg \alpha \rho_\eta $ the one-loop results are stable
and $T^2_{\rm eff} \sim \rho_\eta/|{\bf k}_*|^2$. 
Note that our parameters, $|{\bf k}_*|$ and $\rho_\eta$ are outcome of 
the stage of resonant decay and
depend on the couplings $g$ and $\lambda$ in Eq. (1). For example, we can 
expect that 
$|{\bf k}_*|^2\sim \sqrt{g \lambda}\: M_{\rm Pl}^2$ \re{explosive}. 
In the opposite limit, $ |{\bf k}_*|^4 \ll \alpha \rho_\eta $, we have found
$T^2_{\rm eff} \sim \sqrt{\rho_\eta/\alpha}$. This is smaller then the 
one-loop result by a factor $|{\bf k}_*|^2/\sqrt{\alpha\rho_\eta}$.
This result remains valid when the inflaton decay products $\eta$ and the
order parameter in question correspond to different fields. 
The strength of the symmetry restoration in
a highly non-equilibrium state cannot be traced by the one-loop result in the 
case of sufficiently strong interaction of inflaton decay products. 
This can have important consequences on the issue of symmetry
restoration of various simmetries during the preheating era.

\vspace{36pt}

\centerline{\bf ACKNOWLEDGMENTS}

We are grateful to S. Yu. Khlebnikov for valuable discussions. 
We thank L. Kofman, A. Linde and L. McLerran for comments.
AR is supported by the DOE and NASA under Grant NAG5--2788; IT is  
supported by the DOE grant DE-AC02-76ER01545 at Ohio.

\frenchspacing
\def\prpts#1#2#3{Phys. Reports {\bf #1}, #2 (#3)}
\def\prl#1#2#3{Phys. Rev. Lett. {\bf #1}, #2 (#3)}
\def\prd#1#2#3{Phys. Rev. D {\bf #1}, #2 (#3)}
\def\plb#1#2#3{Phys. Lett. {\bf #1B}, #2 (#3)}
\def\npb#1#2#3{Nucl. Phys. {\bf B#1}, #2 (#3)}
\def\apj#1#2#3{Astrophys. J. {\bf #1}, #2 (#3)}
\def\apjl#1#2#3{Astrophys. J. Lett. {\bf #1}, #2 (#3)}
\begin{picture}(400,50)(0,0)
\put (50,0){\line(350,0){300}}
\end{picture}

\vspace{0.25in}

\def\labelenumi{[\theenumi]}

\begin{enumerate}

\item\label{explosive} L. Kofman, A.D. Linde  and A.A. Starobinsky,
 Phys. Rev. Lett. {\bf 73}, 3195 (1994).

\item\label{linde} A. D. Linde, {\em Particle Physics and Inflationary
Cosmology} (Harwood, Chur, Switzerland, 1990); 
E. W. Kolb and M. S. Turner, {\em The Early Universe} (Addison-Wesley,
Redwood City, California, 1990).


\item\label{st} S. Yu. Khlebnikov and I. I. Tkachev, {\it Classical
decay of inflaton}, hep-ph 9603378.

\item\label{noneq} D. Boyanovsky, D.-S. Lee and A. Singh, Phys. Rev.  
{\bf D48}, 800   (1993);
D.  Shtanov,  J. Traschen and R. Brandenberger,
 Phys. Rev. {\bf D51}, 5438 (1995);
D. Boyanovsky, H.J. de Vega, R. Holman, D.-S. Lee and A. Singh,
 Phys. Rev. {\bf D51}, 4419 (1995);
D. Boyanovsky,  M. D'Attanasio, H.J. de Vega, R. Holman and D.-S.  
Lee, Phys. Rev. {\bf D52}, 6805 (1995);
D. Kaiser, Phys. Rev. {\bf D53} (1996) 1776; D. T. Son,  
hep-ph/9604340.

\item\label{kolb} E.W. Kolb and A. Riotto, FERMILAB-Pub-96-036-A,  
astro-ph 9602095.

\item\label{KLSSR} L. Kofman, A.D. Linde  and A.A. Starobinsky,
 Phys. Rev. Lett. {\bf 76} (1996) 1011.

\item\label{tkachev} I. Tkachev, OSU-TA-21/95 preprint, astro-ph  
9510146, to be published in Phys. Lett. {\bf B}.

\item\label{mono} Ya.B. Zel'dovich and M. Yu. Khlopov,
 Phys. Lett. {\bf B79}, 239 (1978).

\item\label{dw} Ya.B. Zel'dovich, I. Yu. Kobzarev and L. Okun',
 Sov. Phys. JETP {\bf 40}, 1 (1974).

\item\label{tfd} H. Umezawa, H. Matsumoto and M. Tachiki,
 {\it Thermo Field Dynamics and Condensed States}, North Holland,  
1982; P.A. Henning, Phys. Rep. {\bf 253} (1995) 235.

\item\label{mass}Y. Fujimoto and R. Grigjanis, Z. Phys. {\bf C28}  
(1985) 348.

\item\label{pert} T. Bibilashvili and I. Paziashvili, Ann. Phys. (NY)  
{\bf 220} (1992) 134; T. Altherr and D. Seibert, Phys. Lett. {\bf  
B333} (1994) 149;
T. Altherr, Phys. Lett. {\bf B341} (1994) 325; P.F. Badeque, Phys.  
Lett. {\bf 344} (1995) 23.

\item\label{dj} L. Dolan and R. Jackiw, Phys. Rev. {\bf D9} (1974)  
3320.

\item\label{cjt} J.M. Cornwall, R. Jackiw and E. Tomboulis, Phys.  
Rev. {\bf D10} (1974) 2428.

\item\label{chou} See, for instance, K. Chou, B. Hao and L. Yu, Phys.  
Rep. {\bf 118} (1985) 1.

\item\label{limit} R. Jackiw, Phys. Rev. {\bf D9} (1974) 1686.

\end{enumerate}

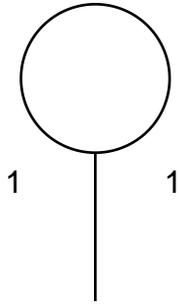
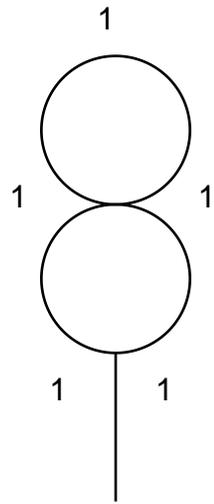
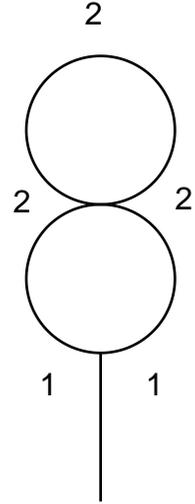
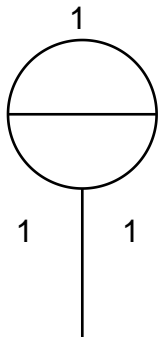
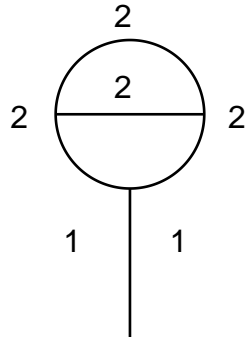
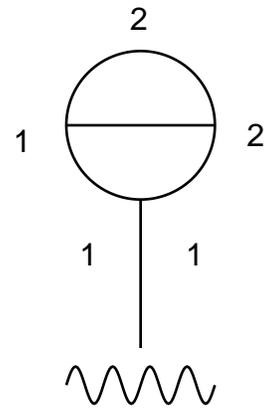
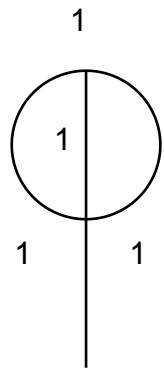
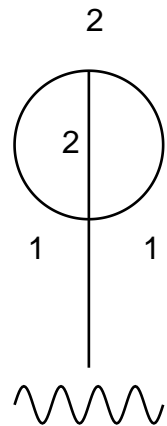
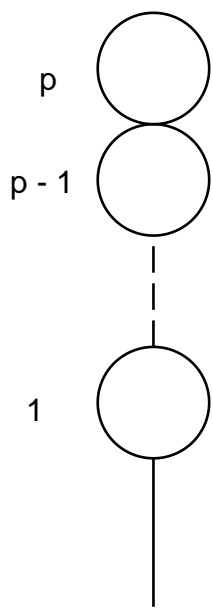
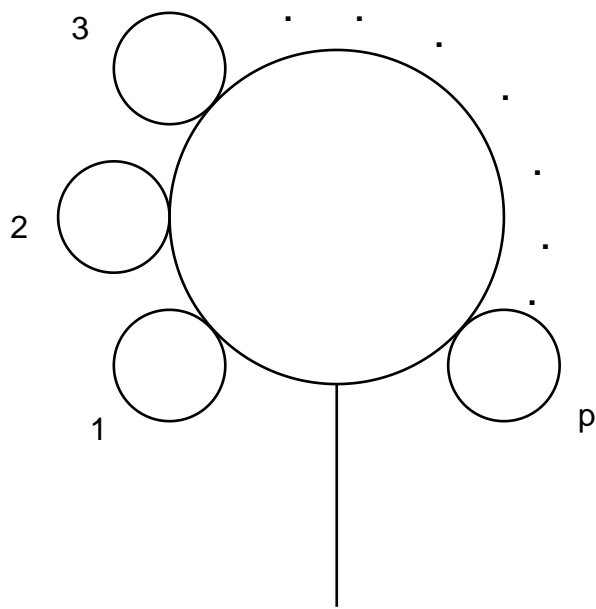
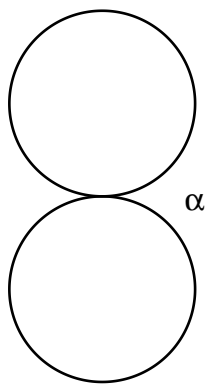
\end{document}